\documentstyle[12pt]{article}

\newcounter{eq}
\newcounter{sc}

\def\overleftrightarrow#1{\vbox{\ialign{##\crcr
 $\leftrightarrow$\crcr\noalign{\kern-1pt\nointerlineskip}
 $\hfil\displaystyle{#1}\hfil$\crcr}}}

\newlength{\minitwocolumn}
\begin{document}

%%%%%%%%%%%%%%%%%%%%%%%%%%%%%%%%%%%%%%%%%%%%%%%%%%%%%%%%%%%%%%%%%%
%%%%%%%%%%%%%%%%%%%%%%%% Title %%%%%%%%%%%%%%%%%%%%%%%%%%%%%%%%%%%
%%%%%%%%%%%%%%%%%%%%%%%%%%%%%%%%%%%%%%%%%%%%%%%%%%%%%%%%%%%%%%%%%%
\begin{flushright}
DPUR/TH/29\\
December, 2011\\
%hep-th/070****\\
\end{flushright}
\vspace{20pt}

%\magnification=\magstep1
\pagestyle{empty}
\baselineskip15pt
%\font\cmssB=cmss17
%\font\cmssS=cmss10

\begin{center}
{\large\bf Subluminal OPERA Neutrinos
\vskip 1mm }

\vspace{20mm}
Ichiro Oda \footnote{E-mail address:\ ioda@phys.u-ryukyu.ac.jp
}

\vspace{5mm}
           Department of Physics, Faculty of Science, University of the 
           Ryukyus,\\
           Nishihara, Okinawa 903-0213, Japan.\\

\end{center}

%\maketitle

\vspace{5mm}
\begin{abstract}
The OPERA collaboration has announced to have observed superluminal neutrinos
with a mean energy 17.5 GeV, but afterward the superluminal interpretation 
of the OPERA results has been refuted theoretically by Cherenkov-like radiation
and pion decay. In a recent work, we have proposed a kinematical resolution to 
this problem. A key idea in our resolution is that the OPERA neutrinos are not superluminal 
but subluminal since they travel faster than the observed speed of light in vacuum 
on the earth while they do slower than the true speed of light in vacuum determining 
the causal structure of events. In this article, we dwell upon our ideas and present some 
concrete models, which realize our ideas, based on spin 0, 1 and 2 bosonic fields.
We also discuss that the principle of invariant speed of light in special relativity
can be replaced with the principle of a universal limiting speed.
\end{abstract}

\newpage
\pagestyle{plain}
\pagenumbering{arabic}
%\setcounter{page}{1}

%%%%%%%%%%%%%%%%%%%%%%%%%%%%%%%%%%%%%%%%%%%%%%%%%%%%%%%%%%%%%%%%%%
%%%%%%%%%%%%%%%%%%%%%%%% Article %%%%%%%%%%%%%%%%%%%%%%%%%%%%%%%%%
%%%%%%%%%%%%%%%%%%%%%%%%%%%%%%%%%%%%%%%%%%%%%%%%%%%%%%%%%%%%%%%%%%

\rm
%%%%%%%%%%%%%%%%%%%%%%%%%%%%%%%%%%%%%%%%%%%%%%%%%%%%%%%%%%%%%%%%%%%%%
%%%%%%%%%%%%%%%%%%%%%%%%%%%%%%   SEC  1    %%%%%%%%%%%%%%%%%%%%%%%%%%
%%%%%%%%%%%%%%%%%%%%%%%%%%%%%%%%%%%%%%%%%%%%%%%%%%%%%%%%%%%%%%%%%%%%%
\section{Introduction}

The OPERA collaboration has recently claimed that it has observed
a superluminal speed of of muon neutrinos \cite{OPERA}. The CNGS
beam of neutrinos with a mean energy 17.5 GeV ranging up to 50 GeV
travels along the baseline distance 730km from CERN to the Gran Sasso laboratory.
The OPERA has found that the neutrinos arrived earlier than expected from 
the speed of light by about 60 nano-seconds. (In the recent second experiment 
using a shorter neutrino beam, the time is about 62 nano-seconds faster than the
speed of light which is within errors of the first experiment.)

An important theoretical challenge is to reconcile the OPERA superluminal
neutrinos with the subluminal neutrinos traveled from SN1987A \cite{Hirata}. 
Average energy of the OPERA neutrinos is a thousand times larger than 
that of the SN1987A neutrinos. This observation prompts some people to 
hit on an idea that only high-energy neutrinos might be superluminal 
whereas low-energy ones are subluminal. The simplest possible approach realizing 
this idea is to change the conventional dispersion relation of a neutrino
by adding the Lorentz-invariant and/or the Lorentz-noninvariant terms, 
thereby making it possible to change the velocity of the neutrino in an energy-dependent
way in order to agree with both the experiments. However, although this resolution 
might be a logical possibility, we think it quite unnatural since both 
the OPERA superluminal neutrinos with about 17 GeV and the SN1987A subluminal ones 
with about 20 MeV are already in the ultra-high energy region compared to its characteristic 
energy scale, which is equal to mass of neutrinos, $m(\nu_e) \le 2.5 eV$ for 
an electric neutrino and $m(\nu_\mu) \le 170 keV$ for a muon neutrino \cite{Nakamura}. 

The other challenges to the OPERA experimental results are the bremsstrahlung 
effect \cite{Cohen} and pion decay \cite{Gonzalez, Bi, Cowsik}. 
The bremsstrahlung effect (or sometimes called Cherenkov-like radiation or Cohen-Glashow
effect) of superluminal neutrinos is a characteristic feature of $\it{superluminal}$ 
neutrinos with weak interaction
such that even if neutrinos do not carry electric charges, the excessive kinetic energy
of the superluminal motion is converted to energy for creating a pair of electron and positron. 
After some calculations, it turns out that on the way from CERN to Gran Sasso, 
the very effect of superluminal propagation of neutrinos would have caused some 
distortions in the beam of neutrinos owing to the bremsstrahlung effect and severely 
depleted the higher-energy neutrinos, thereby making it impossible to observe neutrinos 
with more than 12.5 GeV energy. 
This theoretical result is obviously against the OPERA results where a lot of high-energy 
neutrinos above 12.5 GeV are observed \cite{OPERA}.

More recently, stimulated with the above theoretical observation, ICARUS group has
analyzed their data and found that the neutrino energy distribution of the ICARUS
events in IAr does not have such a distorted energy distribution of beam from CERN. Thus,
the ICARUS group also refutes a superluminal interpretation of the OPERA results on the basis
of the Cohen and Glashow prediction for a weak current analog to Cherenkov radiation \cite{ICARUS}.  

In this article, let us suppose that the OPERA results are correct
even if further experimental scrutiny is surely needed. 
It is then true that a confirmation of the superluminal neutrino motion might require
a radical reconsideration of fundamental principles behind particle physics. 
A successful theory which explains the OPERA results quantitatively as well as
qualitatively should be consistent with very restrictive bounds on the violation
of Lorentz invariance in the sector of charged particles, with the absence of
abnormal dispersion of the neutrino signal from SN1987A, and with the absence of
intensive neutrino decays which are characteristic features for many models
with derivations from the relativistic invariance. 

However, since special relativity has passed many of nontrivial both experimental
and theoretical tests thus far and has a firm foundation, we believe
that special relativity cannot be ruled out by such a single experiment of the speed 
of neutrinos. We therefore wish to conjecture that if the OPERA report is correct, 
it might suggest that the concept of the velocity of light in vacuum must be modified 
to some degree without changing the essential contents of special relativity.\footnote{A similar
idea has been put forward by Nakanishi \cite{Nakanishi}. See also related works 
\cite{Moffat, Li, Gardner}. }

We shall postulate existence of new bosonic degrees of freedom with spins 0, 1 and 2, 
and then assume that these new fields are sourced by the earth, in particular, its energy-momentum
tensor or electro-magnetic current and create a classical background. In contrast to 
previous models, it is assumed that in our models only photons couple to the classical 
background and consequently they propagate on the background via an effective metric.

This article is organized as follows: In the next section, we wish to explain that
the principle of invariant light speed in special relativity can be
replaced with the principle of a universal limiting speed. This limiting
speed is nothing but the speed which appears in various formulae in special relativity.
In Section 3, we will present three models which have two speeds of light because of interactions 
between the gauge field and bosonic fields with spins 0, 1 and 2, which are generalizations
of our previous model \cite{Oda2}.
Section 4 is devoted to discussion.

%%%%%%%%%%%%%%%%%%%%%%%%%%%%%%%%%%%%%%%%%%%%%%%%%%%%%%%%%%%%%%%%%%%%%
%%%%%%%%%%%%%%%%%%%%%%%%%%%%%%   SEC  2    %%%%%%%%%%%%%%%%%%%%%%%%%%
%%%%%%%%%%%%%%%%%%%%%%%%%%%%%%%%%%%%%%%%%%%%%%%%%%%%%%%%%%%%%%%%%%%%%
\section{Review of special relativity and our resolution to Cohen-Glashow effect}

In this section, we review special relativity \cite{Einstein}, 
in particular, the principle of invariant speed of light and present 
our resolution to Cohen-Glashow effect of superluminal neutrinos.

The special theory of relativity by Einstein, which is often called 
$\it{special \ relativity}$, has become a commonplace in physics, as
taken for granted as Newton's laws of classical mechanics and Maxwell
equations of electrodynamics. It is well known that Einstein's special
relativity is based on two fundamental principles, those are, the principle
of special relativity and the principle of invariant speed of light.

The former principle says that all physical equations must be invariant under
Lorentz transformations and it is a universal principle which holds for every
physical phenomenon. On the other hand, the latter principle, which implies that
the speed of light is finite and independent of the motion of its source, 
supposes existence of light from the outset. Here note that light is a kind of 
electromagnetic waves whose existence is guaranteed by electromagnetics. 
In this context, it seems to be a bit strange to accept the principle of invariant light 
speed as one of the fundamental principles in special relativity since this principle 
attaches too much importance to electrodynamics. Special relativity as well as 
general relativity are theories of space and time so they should be defined prior to 
other branches of physics such as electrodynamics.

It is therefore natural to imagine that we might be able to construct special relativity 
without relying on the existence of light coming from electrodynamics.
Indeed, it was known quite some time ago that we can make special relativity in a such
way that it is not based on the principle of invariant light speed as one of the fundamental 
principles \cite{Ignatowsky, Frank, Mermin, Jackson}.  
In such an approach, the principle of invariant speed of light is replaced 
with the principle of a universal limiting speed, which means that in every inertial 
frame, there is a finite universal limiting speed $C$ for all physical entities.
In passing we remark that the existence of a universal limiting speed could be rephrased that
there is a upper limit of speed in the propagation of information.

Now we wish to show explicitly that both the principle of special relativity and
that of a universal limiting speed produce the Lorentz transformation and the
composition law of velocities in special relativity. To do that, let $I$ and $I'$
be two inertial, non-accelerating reference frames in such a way that $I'$ moves in the
direction of the $x$ axis with a relative velocity $v$ compared to $I$. (For simplicity
we omit $y$ and $z$ coordinates.)  Then, it is easy to see that the principle of
special relativity together with isotropy and homogeneity of the Minkowski space-time
leads to the most general transformation between the coordinates $(t, x)$ in $I$ and
$(t', x')$ in $I'$
%**   Transf %%%%%%%%%%%%%%%%%%%%%%%%%%%%%%%%%%%%%%%%%%%%%%%%%%%%%%%%%
\begin{eqnarray}
t' &=&  A(v^2) t - B(v^2) v x, \nonumber\\
x' &=& D(v^2) ( x - v t ),
\label{Transf}
\end{eqnarray}
%%%%%%%%%%%%%%%%%%%%%%%%%%%%%%%%%%%%%%%%%%%%%%%%%%%%%%%%%%%%%%%%%%%
and its inverse transformation 
%**   I-Transf %%%%%%%%%%%%%%%%%%%%%%%%%%%%%%%%%%%%%%%%%%%%%%%%%%%%%%%%%
\begin{eqnarray}
t &=&  A(v^2) t' + B(v^2) v x', \nonumber\\
x &=& D(v^2) ( x' + v t' ),
\label{I-Transf}
\end{eqnarray}
%%%%%%%%%%%%%%%%%%%%%%%%%%%%%%%%%%%%%%%%%%%%%%%%%%%%%%%%%%%%%%%%%%%
where $A$, $B$ and $C$ are functions of only $v^2$, and note that the common factor $D$ 
is needed from the definition of inertial frames in the relative motion.

With this most general transformation, the consistency condition between (\ref{Transf})
and (\ref{I-Transf}) yields relations
%**   Consistency %%%%%%%%%%%%%%%%%%%%%%%%%%%%%%%%%%%%%%%%%%%%%%%%%%%%%%%%%
\begin{eqnarray}
A = D, \quad  ( A - v^2 B ) D = 1.
\label{Consistency}
\end{eqnarray}
%%%%%%%%%%%%%%%%%%%%%%%%%%%%%%%%%%%%%%%%%%%%%%%%%%%%%%%%%%%%%%%%%%%
If a physical object has a velocity $V'$ in the reference frame $I'$, the corresponding 
velocity $V$ in the reference frame $I$ reads
%**   V %%%%%%%%%%%%%%%%%%%%%%%%%%%%%%%%%%%%%%%%%%%%%%%%%%%%%%%%%
\begin{eqnarray}
V \equiv \frac{dx}{dt} = \frac{\frac{dx}{dt'}} {\frac{dt}{dt'}} 
= \frac{V' + v}{1 + v V' \frac{B}{A}},
\label{V}
\end{eqnarray}
%%%%%%%%%%%%%%%%%%%%%%%%%%%%%%%%%%%%%%%%%%%%%%%%%%%%%%%%%%%%%%%%%%%
where Eq. (\ref{Consistency}) is used.
Then, the principle of a universal limiting speed $C$ allows us to choose
%**   C %%%%%%%%%%%%%%%%%%%%%%%%%%%%%%%%%%%%%%%%%%%%%%%%%%%%%%%%%
\begin{eqnarray}
\frac{B}{A} = \frac{1}{C^2}.
\label{C}
\end{eqnarray}
%%%%%%%%%%%%%%%%%%%%%%%%%%%%%%%%%%%%%%%%%%%%%%%%%%%%%%%%%%%%%%%%%%%
As a result, we arrive at the Lorentz transformation
%**   Lorentz-transf %%%%%%%%%%%%%%%%%%%%%%%%%%%%%%%%%%%%%%%%%%%%%%%%%%%%%%%%%
\begin{eqnarray}
t' &=&  \frac{t - \frac{v}{C^2} t}{\sqrt{1 - (\frac{v}{C})^2}}, \nonumber\\
x' &=& \frac{x - v t}{\sqrt{1 - (\frac{v}{C})^2}} ,
\label{Lorentz-transf}
\end{eqnarray}
%%%%%%%%%%%%%%%%%%%%%%%%%%%%%%%%%%%%%%%%%%%%%%%%%%%%%%%%%%%%%%%%%%%
and the composition law of velocities
%**   Composition law %%%%%%%%%%%%%%%%%%%%%%%%%%%%%%%%%%%%%%%%%%%%%%%%%%%%%%%%%
\begin{eqnarray}
V =  \frac{V' + v}{1 + \frac{v V'}{C^2}}.
\label{Composition law}
\end{eqnarray}
%%%%%%%%%%%%%%%%%%%%%%%%%%%%%%%%%%%%%%%%%%%%%%%%%%%%%%%%%%%%%%%%%%%

Moreover, it has been shown by Terletskii \cite{Terletskii} that by
considering three inertial reference frames $I(t, x)$, $I'(t', x')$ and $I''(t'', x'')$,
and the group property of the two transformations $(t, x) \rightarrow (t', x')
\rightarrow (t'', x'')$ and $(t, x) \rightarrow (t'', x'')$ directly, 
$C$ is a universal constant with the dimension of a speed. In addition to it,
one can show within the framework of this approach that massive particles cannot
move with the velocity of the universal limiting speed $C$ whereas massless particles
can do so. As for light, QED (Quantum Electro-Dynamics) requires photons, 
which are quanta of light, to be massless and consequently 
light can propagate with the velocity of the
upper limit $C$.\footnote{We identify the velocity of a photon with that of
light.} Put differently, $C$ is not in essence the velocity of light 
but light can move at the velocity $C$ since its quanta 'photons' happen to be 
massless through the quantization of Maxwell's electrodynamics. Incidentally,
gravitational waves in general relativity can propagate at the velocity $C$ as well
since their quanta 'gravitons' are massless.

In this way, by starting with both the principle of special relativity and the
principle of a universal limiting speed without referring to the existence
of light, one can reproduce the Lorentz transformation and the composition law
of velocities in Einstein's special relativity. An important point in this approach is 
that the universal limiting speed $C$ is to be determined only through experiment. 
It is this experiment that we are interested in connection with a resolution
to the Cohen-Glashow effect of superluminal neutrinos.

Before delving into our resolution \cite{Oda2}, let us review the Cohen-Glashow effect 
briefly \cite{Cohen}.
The basic observation behind the Cohen-Glashow effect is that a neutrino 
moving faster than the velocity of light loses its kinetic energy by emitting
something via weak interaction even if it does not possess electric charges.
The most dominant decay process of a muon neutrino, which mainly constitutes the
OPERA beam, is given by $\nu_\mu \rightarrow \nu_\mu + e^+ + e^-$. It turns out
that after some calculations the terminal energy of a neutrino detected at Gran
Sasso is about 12.5GeV, so few neutrinos with energies in excess of 12.5GeV reach
the detector. Unfortunately, since the OPERA detector observes neutrinos with the mean
energy of 17.5GeV ranging up to 50GeV, the Cohen-Glashow effect rules out an
interpretation of the OPERA neutrinos being superluminal \cite{Cohen}. Furthermore, the ICARUS
group also reaches the same conclusion as Cohen and Glashow by reanalyzing their data \cite{ICARUS}.

Now we are ready to present our resolution to the Cohen-Glashow effect.
First, let us note that the Cohen-Glashow effect is a peculiar feature
of superluminal particles. It is worthwhile to recall that special relativity 
never forbids existence of superluminal particles but only prohibits superluminal particles
from becoming subluminal particles, and vice versa because the proper Lorentz transformation
does not connect superluminal motion with subluminal one. 
Thus, if a particle is superluminal, we cannot take the rest frame for the particle 
since its minimum speed must be more than a universal limiting speed. 
On the other hand, if a particle is subluminal, by performing a suitable Lorentz transformation
we can always take the rest frame for this particle, 
thereby kinematically forbidding the Cherenkov-like process 
of superluminal neutrinos to occur. Hence, our resolution to the Cohen-Glashow effect
amounts to saying that the OPERA neutrinos are not superluminal but subluminal in order
to avoid the Cohen-Glashow effect \cite{Oda2}. Then, a natural question arises whether our resolution
is against the interpretation of superluminal neutrinos by OPERA group or not? 

In order to answer this question, let us note that the causal structure, in other words, 
the fact whether a particle is superluminal, light-like or subluminal, is determined by 
the universal limiting speed $C$, which should be measured by experiment as mentioned above.
There are different methods to determine the value of $C$ experimentally.
One natural way is to measure the actual speed of light in vacuum, which
can be done in various astronomical and earth-based setups. Here we wish to
insist that astronomical measurement should be done in outer space far from stars 
and planets in order to avoid the influence of dark matters.
The experiments for measuring the speed of light in outer space have been thus far
done by using the earth and various planets such as the sun and the moon in the solar system.
But these experiments cannot help receiving the influence of dark matters since
dark matters have a tendency to gather near massive objects such as stars and planets 
via a gravitational interaction. The universal limiting speed $C$ at hand must be measured 
in a setup for which there is no coupling to dark matters.

As is well known, the phase velocity $v$ at which light propagates in a medium such as
water or air, is reduced by the refractive index $n$ of the medium as
$v = \frac{C}{n} < C$. If there are some undetectable media on the earth
like dark matters, it is plausible that the observed speed $c$ of light
on the earth also becomes smaller than the universal limiting speed $C$, that is 
$c < C$. Then, it is reasonable to conjecture that the observed speed $v_\nu$ 
of the OPERA neutrinos is larger than the observed speed $c$ of light on the earth 
while it is smaller than the universal limiting speed $C$\footnote{Of course,
when there is no coupling between photons and dark matters, the observed
speed of light coincides with the universal limiting speed $C$.}
%**   Conjecture %%%%%%%%%%%%%%%%%%%%%%%%%%%%%%%%%%%%%%%%%%%%%%%%%%%%%%%%%
\begin{eqnarray}
c < v_\nu < C.
\label{Conjecture}
\end{eqnarray}
%%%%%%%%%%%%%%%%%%%%%%%%%%%%%%%%%%%%%%%%%%%%%%%%%%%%%%%%%%%%%%%%%%%
With this conjecture, the OPERA neutrinos should be regarded as subluminal neutrinos 
since the causal structure of events are now defined with respect to the universal 
limiting speed $C$ as mentioned before.
 
In other words, what we wish to present as a resolution to the Cohen-Glashow
effect is the following: the OPERA neutrinos might be superluminal ($c < v_\nu$) 
if one compares the velocity of a neutrino with the observed velocity $c$ of light 
on the earth, but since superluminality or subluminality must be determined 
on the basis of the universal limiting speed $C$, the OPERA neutrinos are 
actually subluminal ($v_\nu < C$) in comparison with the true velocity $C$.

%%%%%%%%%%%%%%%%%%%%%%%%%%%%%%%%%%%%%%%%%%%%%%%%%%%%%%%%%%%%%%%%%%%%%
%%%%%%%%%%%%%%%%%%%%%%%%%%%%%%   SEC  3    %%%%%%%%%%%%%%%%%%%%%%%%%%
%%%%%%%%%%%%%%%%%%%%%%%%%%%%%%%%%%%%%%%%%%%%%%%%%%%%%%%%%%%%%%%%%%%%%
\section{Models with two velocities of light}

Now let us note that with our conjecture mentioned above, 
the problem of the bremsstrahlung effect is converted to a different problem, 
which can be stated as follows: 
"Can we construct a physically plausible model which explains how a universal 
limiting speed $C$ could be changed to the observed speed $c$ of light 
on the earth by the influence of dark matters?".\footnote{See Ref. \cite{Ellis}
on the meaning of the speed of light and the varying speed of light theories.}

In our previous work \cite{Oda2}\footnote{See also related works 
\cite{Dvali, Iorio, Kehagias, Wang, Saridakis, Oda1, Hebecker}.}, 
we have presented such a model with the observed velocity of light by using 
a symmetric tensor field of spin 2. In this section, we will construct 
three types of models with the observed velocity of light by using three 
kinds of bosonic degrees of freedom, which are scalar, vector and symmetric tensor fields.
These bosonic fields are sourced by the earth and create a classical background, to which
a gauge field describing photons couples. The photons then propagate through an
effective metric and consequently the universal limiting speed $C$ will be slightly 
reduced to become the observed speed of light $c$ on the earth.  Note that these bosonic 
fields could be regarded as parts of many candidates for dark matters and behave as if they were media
such as water or air.

\subsection{Scalar field} 

Let us start with the case of a scalar field whose effective Lagrangian density is 
of form:\footnote{We make use of a flat metric $\eta_{\mu\nu} = diag ( -1, +1, +1, +1)$ 
for raising or lowering indices. Moreover, we adopt the Planck units $C = \hbar = G = 1$
by which all quantities become dimensionless multiples of the Planck length $L_{Pl}
\equiv \left( \frac{G \hbar}{C^3} \right)^{\frac{1}{2}}$.}
%**   S-Lag %%%%%%%%%%%%%%%%%%%%%%%%%%%%%%%%%%%%%%%%%%%%%%%%%%%%%%%%%
\begin{eqnarray}
{\cal{L}} = - \frac{1}{4} F_{\mu\nu} F^{\mu\nu} 
+ \frac{1}{2 M_*^4} \partial^\nu \Pi \partial^\alpha \Pi F_{\mu\nu} F^\mu \ _\alpha
- \frac{1}{2} \partial_\mu \Pi \partial^\mu \Pi - \frac{m^2}{2} \Pi^2
+ \frac{4 \pi}{M} \Pi T,
\label{S-Lag}
\end{eqnarray}
%%%%%%%%%%%%%%%%%%%%%%%%%%%%%%%%%%%%%%%%%%%%%%%%%%%%%%%%%%%%%%%%%%%
where $M_*$ is a mass scale which controls the strength of a coupling between
the scalar field $\Pi$ and the abelian gauge field $A_\mu$. And $M$ is another
mass scale setting the strength of a coupling between the scalar field and
the trace of the energy-momentum tensor which describes an effective energy-momentum
except the gauge field of the earth.
The gauge field strength $F_{\mu\nu}$ is defined by $F_{\mu\nu} = \partial_\mu A_\nu
- \partial_\nu A_\mu$ as usual.  Note that the second term is non-renormalizable
so this Lagrangian density makes sense for energies less than $M_*$, beyond
which the theory enters in the strong coupling region.
Note that in the absence of the $\Pi$ field, the photon travels at the universal
limiting speed $C$ since the gauge field satisfies the conventional
Maxwell equations. 

Now it is easy to rewrite (\ref{S-Lag}) as
%**   S-Lag 2 %%%%%%%%%%%%%%%%%%%%%%%%%%%%%%%%%%%%%%%%%%%%%%%%%%%%%%%%%
\begin{eqnarray}
{\cal{L}} = - \frac{1}{4} ( \eta^{\mu\alpha} - \frac{1}{M_*^4} \partial^\mu \Pi 
\partial^\alpha \Pi ) ( \eta^{\nu\beta} - \frac{1}{M_*^4} \partial^\nu \Pi 
\partial^\beta \Pi ) F_{\mu\nu} F_{\alpha\beta}
- \frac{1}{2} \partial_\mu \Pi \partial^\mu \Pi - \frac{m^2}{2} \Pi^2
+ \frac{4 \pi}{M} \Pi T,
\label{S-Lag 2}
\end{eqnarray}
%%%%%%%%%%%%%%%%%%%%%%%%%%%%%%%%%%%%%%%%%%%%%%%%%%%%%%%%%%%%%%%%%%%
thereby making it possible to read out an effective metric
%**   S-metric %%%%%%%%%%%%%%%%%%%%%%%%%%%%%%%%%%%%%%%%%%%%%%%%%%%%%%%%%
\begin{eqnarray}
g^{\mu\nu}_{(A)} = \eta^{\mu\nu} - \frac{1}{M_*^4} \partial^\mu \Pi
\partial^\nu \Pi, 
\label{S-metric}
\end{eqnarray}
%%%%%%%%%%%%%%%%%%%%%%%%%%%%%%%%%%%%%%%%%%%%%%%%%%%%%%%%%%%%%%%%%%%
on which the photon propagates. 
 
Next, let us derive the equation of motion for the scalar field $\Pi$
%**   S-Eq %%%%%%%%%%%%%%%%%%%%%%%%%%%%%%%%%%%%%%%%%%%%%%%%%%%%%%%%%
\begin{eqnarray}
( \Box - m^2 ) \Pi = - \frac{4 \pi}{M} T 
+ \frac{1}{M_*^4} \partial^\alpha ( \partial^\nu \Pi 
F_{\mu\nu} F^\mu \ _\alpha ). 
\label{S-Eq}
\end{eqnarray}
%%%%%%%%%%%%%%%%%%%%%%%%%%%%%%%%%%%%%%%%%%%%%%%%%%%%%%%%%%%%%%%%%%%
For the range of energies and distances of our interests, the linearized analysis
is fully sufficient and reliable, so we confine ourselves to the linearized
equation of motion of Eq. (\ref{S-Eq})
%**   S-L-Eq %%%%%%%%%%%%%%%%%%%%%%%%%%%%%%%%%%%%%%%%%%%%%%%%%%%%%%%%%
\begin{eqnarray}
( \Box - m^2 ) \Pi = - \frac{4 \pi}{M} T. 
\label{S-L-Eq}
\end{eqnarray}
%%%%%%%%%%%%%%%%%%%%%%%%%%%%%%%%%%%%%%%%%%%%%%%%%%%%%%%%%%%%%%%%%%%
Here $T_{\mu\nu}$ is taken as a non-relativistic, static and spherically symmetric
source of the earth's mass $M_\oplus$
%**   S-T %%%%%%%%%%%%%%%%%%%%%%%%%%%%%%%%%%%%%%%%%%%%%%%%%%%%%%%%%
\begin{eqnarray}
T_{00} = M_\oplus \delta^3(r),
\label{S-T}
\end{eqnarray}
%%%%%%%%%%%%%%%%%%%%%%%%%%%%%%%%%%%%%%%%%%%%%%%%%%%%%%%%%%%%%%%%%%%
and the other components of $T_{\mu\nu}$ are vanishing. Then, the trace
of $T_{\mu\nu}$ is given by
%**   S-Trace %%%%%%%%%%%%%%%%%%%%%%%%%%%%%%%%%%%%%%%%%%%%%%%%%%%%%%%%%
\begin{eqnarray}
T \equiv \eta^{\mu\nu} T_{\mu\nu} = - T_{00} = - M_\oplus \delta^3(r).
\label{S-Trace}
\end{eqnarray}
%%%%%%%%%%%%%%%%%%%%%%%%%%%%%%%%%%%%%%%%%%%%%%%%%%%%%%%%%%%%%%%%%%%
Thus, for the static configuration, Eq. (\ref{S-L-Eq}) reads
%**   S-L-Eq 2 %%%%%%%%%%%%%%%%%%%%%%%%%%%%%%%%%%%%%%%%%%%%%%%%%%%%%%%%%
\begin{eqnarray}
( \Delta - m^2 ) \Pi = 4 \pi \frac{M_\oplus}{M} \delta^3(r), 
\label{S-L-Eq 2}
\end{eqnarray}
%%%%%%%%%%%%%%%%%%%%%%%%%%%%%%%%%%%%%%%%%%%%%%%%%%%%%%%%%%%%%%%%%%%
so the solution takes the form
%**   S-L-Sol %%%%%%%%%%%%%%%%%%%%%%%%%%%%%%%%%%%%%%%%%%%%%%%%%%%%%%%%%
\begin{eqnarray}
\Pi = - \frac{M_\oplus}{M} \frac{1}{r} e^{-mr}. 
\label{S-L-Sol}
\end{eqnarray}
%%%%%%%%%%%%%%%%%%%%%%%%%%%%%%%%%%%%%%%%%%%%%%%%%%%%%%%%%%%%%%%%%%%
We postulate that the Compton wave-length of the $\Pi$ field is the order
of planetary distances, $\frac{1}{m} \gg r$, so we effectively set $m = 0$. Then, we obtain
%**   S-L-Sol 2 %%%%%%%%%%%%%%%%%%%%%%%%%%%%%%%%%%%%%%%%%%%%%%%%%%%%%%%%%
\begin{eqnarray}
\Pi = - \frac{M_\oplus}{M} \frac{1}{r}. 
\label{S-L-Sol 2}
\end{eqnarray}
%%%%%%%%%%%%%%%%%%%%%%%%%%%%%%%%%%%%%%%%%%%%%%%%%%%%%%%%%%%%%%%%%%%

With this configuration, an effective space-time on which the photon propagates 
has the line element in the spherically symmetric coordinates
%**   Line-S %%%%%%%%%%%%%%%%%%%%%%%%%%%%%%%%%%%%%%%%%%%%%%%%%%%%%%%%%
\begin{eqnarray}
ds^2 \equiv g_{(A) \mu\nu} d x^\mu d x^\nu 
= - dt^2 + \frac{1}{1 - \frac{l^4}{r^4}} dr^2 + r^2 d \Omega^2_2,
\label{Line-S}
\end{eqnarray}
%%%%%%%%%%%%%%%%%%%%%%%%%%%%%%%%%%%%%%%%%%%%%%%%%%%%%%%%%%%%%%%%%%%
where $d \Omega^2_2 \equiv d \theta^2 + \sin^2 \theta d \phi^2$ and we have
defined $l \equiv \frac{1}{M_*} \sqrt{\frac{M_\oplus}{M}}$.

Whenever one obtains a prediction from general relativity the question always arises 
or should arise whether the result obtained really refers to an objective physical measurement 
or whether it has folded into arbitrary subjective elements dependent on our choice
of coordinate system \cite{Weinberg}. In the case at hand, one should ask oneself
what the predictive change in the velocity of the photon really has to do with the positions 
of observed places. In fact, it has been already pointed out that the quantities defined 
in general relativity as an $\it{average}$ velocity of massless particles traveling between 
two distant points can either sub- or superluminal depending on the position of the
observer and the form of the trajectory in the gravitational field \cite{Lust}. Thus, one has to
consider what velocity is most suitable in the present physical setting. 

In this article, we shall adopt a definition of an effective local velocity found by 
Einstein \cite{Einstein 2}.
Usually, in calculating the precession of the perihelia of the mercury and the
bending of light, the Schwarzschild solution to Einstein's equations is utilized,
but Einstein himself has made use of the weak-field approximation of a
gravitational field in the isotropic coordinates. Thus, following Einstein, we
will look for the isotropic coordinates corresponding to the line element (\ref{Line-S})
as follows:
%**   Line-S 2 %%%%%%%%%%%%%%%%%%%%%%%%%%%%%%%%%%%%%%%%%%%%%%%%%%%%%%%%%
\begin{eqnarray}
ds^2 &=& - dt^2 + \frac{1}{1 - \frac{l^4}{r^4}} dr^2 + r^2 d \Omega^2_2  \nonumber\\
&=& - dt^2 + A(\bar r)^2 (d \bar r^2 + \bar r^2 d \Omega^2_2) \nonumber\\
&=& - dt^2 + A(\bar r)^2 (d \bar x^2 + d \bar y^2 + d \bar z^2),
\label{Line-S 2}
\end{eqnarray}
%%%%%%%%%%%%%%%%%%%%%%%%%%%%%%%%%%%%%%%%%%%%%%%%%%%%%%%%%%%%%%%%%%%
where $\bar r^2 = \bar x^2 + \bar y^2 + \bar z^2$. Then, it turns out that
$A(\bar r)$ is given by
%**   S-A %%%%%%%%%%%%%%%%%%%%%%%%%%%%%%%%%%%%%%%%%%%%%%%%%%%%%%%%%
\begin{eqnarray}
A(\bar r) = \frac{r}{\bar r} = \frac{l}{\sqrt{1 + \sqrt{1 - \frac{l^4}{r^4}}}}.
\label{S-A}
\end{eqnarray}
%%%%%%%%%%%%%%%%%%%%%%%%%%%%%%%%%%%%%%%%%%%%%%%%%%%%%%%%%%%%%%%%%%%

Then, an effective velocity $c(r)$ of a photon is defined as
%**   S-velocity %%%%%%%%%%%%%%%%%%%%%%%%%%%%%%%%%%%%%%%%%%%%%%%%%%%%%%%%%
\begin{eqnarray}
c(r) \equiv \sqrt{ \frac{l^2}{2} \left[ \left(\frac{d \bar x}{d t}\right)^2 +  
\left(\frac{d \bar y}{d t}\right)^2 + \left(\frac{d \bar z}{d t}\right)^2 \right] },
\label{S-velocity}
\end{eqnarray}
%%%%%%%%%%%%%%%%%%%%%%%%%%%%%%%%%%%%%%%%%%%%%%%%%%%%%%%%%%%%%%%%%%%
where an overall factor $\frac{l^2}{2}$ is introduced for normalization.
With this definition, the observed velocity of light in vacuum on the earth becomes
%**   S-velocity 2 %%%%%%%%%%%%%%%%%%%%%%%%%%%%%%%%%%%%%%%%%%%%%%%%%%%%%%%%%
\begin{eqnarray}
c(r) \equiv \frac{C}{\sqrt{2}} \left( 1 + \sqrt{1 - \frac{l^4}{r^4}} \right)^{\frac{1}{2}},
\label{S-velocity 2}
\end{eqnarray}
%%%%%%%%%%%%%%%%%%%%%%%%%%%%%%%%%%%%%%%%%%%%%%%%%%%%%%%%%%%%%%%%%%%
where for conveniece the universal limiting velocity $C$ is recovered.
Note that at the spatial infinity $r \rightarrow \infty$, this observed velocity of light
coincides with the universal limiting velocity $C$, that is, 
%**   S-velocity 3 %%%%%%%%%%%%%%%%%%%%%%%%%%%%%%%%%%%%%%%%%%%%%%%%%%%%%%%%%
\begin{eqnarray}
\lim_{r \to \infty} c(r) = C.
\label{S-velocity 3}
\end{eqnarray}
%%%%%%%%%%%%%%%%%%%%%%%%%%%%%%%%%%%%%%%%%%%%%%%%%%%%%%%%%%%%%%%%%%%
Furthermore, note that the difference between $C$ and $c(r)$ is positive-definite
as required from our conjecture
%**   S-velocity 4 %%%%%%%%%%%%%%%%%%%%%%%%%%%%%%%%%%%%%%%%%%%%%%%%%%%%%%%%%
\begin{eqnarray}
C - c(r) = \frac{C}{8} \frac{l^4}{r^4} > 0,
\label{S-velocity 4}
\end{eqnarray}
%%%%%%%%%%%%%%%%%%%%%%%%%%%%%%%%%%%%%%%%%%%%%%%%%%%%%%%%%%%%%%%%%%%
which holds when $\frac{C}{8} \frac{l^4}{r^4} \ll 1$.

On the earth, the observed velocity of light is given by $c(r)$ when the photons
are located at the place whose distance in the radial direction from the center 
of the earth is $r$. Since it is assumed that the speed of a neutrino, denoted as $v_\nu$, 
takes a definite value and is smaller than the universal limiting speed $C$ but larger 
than the observed speed $c(r)$ 
%**   Velocity ansatz %%%%%%%%%%%%%%%%%%%%%%%%%%%%%%%%%%%%%%%%%%%%%%%%%%%%%%%%%
\begin{eqnarray}
c(r) < v_\nu < C,
\label{Velocity ansatz}
\end{eqnarray}
%%%%%%%%%%%%%%%%%%%%%%%%%%%%%%%%%%%%%%%%%%%%%%%%%%%%%%%%%%%%%%%%%%%
we have a superluminal neutrino for the velocity $c(r)$ as observed in the
OPERA experiment
%**   S-Delta(r) %%%%%%%%%%%%%%%%%%%%%%%%%%%%%%%%%%%%%%%%%%%%%%%%%%%%%%%%%
\begin{eqnarray}
\beta(r) \equiv \frac{v_\nu - c(r)}{c(r)} > 0,
\label{S-Delta(r)}
\end{eqnarray}
%%%%%%%%%%%%%%%%%%%%%%%%%%%%%%%%%%%%%%%%%%%%%%%%%%%%%%%%%%%%%%%%%%%
whereas we have a subluminal neutrino for the universal limiting speed $C$
%**   S-Delta %%%%%%%%%%%%%%%%%%%%%%%%%%%%%%%%%%%%%%%%%%%%%%%%%%%%%%%%%
\begin{eqnarray}
\beta \equiv \frac{v_\nu - C}{C} < 0.
\label{S-Delta}
\end{eqnarray}
%%%%%%%%%%%%%%%%%%%%%%%%%%%%%%%%%%%%%%%%%%%%%%%%%%%%%%%%%%%%%%%%%%%

Recalling that the property of superluminality or subluminality of neutrinos
is defined by using the universal limiting velocity $C$,
the OPERA neutrinos are actually not superluminal but subluminal as emphasized before.
Hence, we do not have Cherenkov-like radiation or the Cohen-Glashow effect 
for the OPERA neutrinos at all since we can always take the rest frame 
for the subluminal neutrino. 
This is our resolution to the problem of the bremsstrahlung effect of 
the OPERA neutrinos.
It is worthwhile to notice that our solution is purely kinematical as 
desired. 

Let us show that the results of OPERA and SN1987A give us useful information
on two mass scales $M_*$ and $M$. First, let us note that the OPERA result yields
a condition for the dimensionless quantity $\beta(r)$
%**   S-Beta %%%%%%%%%%%%%%%%%%%%%%%%%%%%%%%%%%%%%%%%%%%%%%%%%%%%%%%%%
\begin{eqnarray}
\beta(R_\oplus) \equiv  \frac{v_\nu - c(R_\oplus)}{c(R_\oplus)} \approx 2.5 \times 10^{-5},
\label{S-Beta}
\end{eqnarray}
%%%%%%%%%%%%%%%%%%%%%%%%%%%%%%%%%%%%%%%%%%%%%%%%%%%%%%%%%%%%%%%%%%%
where $R_\oplus$ is the radius of the earth and takes the value $R_\oplus = 6.4 
\times 10^8 cm$.

Next, let us recall the fact that neutrinos from SN1987A to the earth travel
at almost the same velocity as the velocity of light in vacuum, so we would have a relation
%**   SN1987A %%%%%%%%%%%%%%%%%%%%%%%%%%%%%%%%%%%%%%%%%%%%%%%%%%%%%%%%%
\begin{eqnarray}
v_\nu \approx C.
\label{SN1987A}
\end{eqnarray}
%%%%%%%%%%%%%%%%%%%%%%%%%%%%%%%%%%%%%%%%%%%%%%%%%%%%%%%%%%%%%%%%%%%
With the help of this relation (\ref{SN1987A}), Eq. (\ref{S-Beta}) can be rewritten as
%**   S-Beta 2 %%%%%%%%%%%%%%%%%%%%%%%%%%%%%%%%%%%%%%%%%%%%%%%%%%%%%%%%%
\begin{eqnarray}
\beta(R_\oplus) \approx  \frac{1}{8} \frac{l^4}{R_\oplus^4} 
= \frac{1}{8 M_*^4} \left(\frac{M_\oplus}{M}\right)^2 \frac{1}{R_\oplus^4}
\approx 2.5 \times 10^{-5}.
\label{S-Beta 2}
\end{eqnarray}
%%%%%%%%%%%%%%%%%%%%%%%%%%%%%%%%%%%%%%%%%%%%%%%%%%%%%%%%%%%%%%%%%%%
Introducing two length scales $L_* = \frac{1}{M_*}$ and $L = \frac{1}{M}$,
we have an equation
%**   S-L %%%%%%%%%%%%%%%%%%%%%%%%%%%%%%%%%%%%%%%%%%%%%%%%%%%%%%%%%
\begin{eqnarray}
L_*^4 L^2 \approx  8 \times 10^{-4} L_{Pl}^4 \frac{R_\oplus^4}{R_{\oplus SS}^2}
\approx 10^{-99} (cm)^6,
\label{S-L}
\end{eqnarray}
%%%%%%%%%%%%%%%%%%%%%%%%%%%%%%%%%%%%%%%%%%%%%%%%%%%%%%%%%%%%%%%%%%%
where $L_{Pl}, R_{\oplus SS}$ are respectively the Planck length and the Schwarzschild
radius of the earth, and are explicitly given by $L_{Pl} = 1.6 \times 10^{-33} cm,
R_{\oplus SS} = 0.89 cm$. 

For instance, if $L_* \approx L$, Eq. (\ref{S-L}) gives us 
%**   S-Exam  %%%%%%%%%%%%%%%%%%%%%%%%%%%%%%%%%%%%%%%%%%%%%%%%%%%%%%%%%
\begin{eqnarray}
L_* \approx L \approx 10^{-17} cm,  \quad M_* \approx M \approx 1 TeV.
\label{S-Exam}
\end{eqnarray}
%%%%%%%%%%%%%%%%%%%%%%%%%%%%%%%%%%%%%%%%%%%%%%%%%%%%%%%%%%%%%%%%%%%
It is of interest that in this instance the mass scale $M_*$ is approximately 
beyond the energy scale of Standard Model. Recall that in the energy region 
above $M_*$, the coupling between the gauge field and the scalar field becomes 
so strong that our effective theory is expected to be replaced to a unknown,
sensible and renormalizable UV-completed theory.
 
At this stage, it is tempting to identify the $\Pi$ scalar field with part
of many candidates for dark matters. Actually, this identification seems to be
consistent with the results of OPERA and SN1987A at the same time by
the following reasoning: In general, dark matters are expected to be
trapped in the vicinity of massive objects such as stars and the earth 
compared to empty regions of outer space owing to a gravitational interaction.
The reason why neutrinos from SN1987A to the earth traveled at almost the same 
velocity as the velocity of light in vacuum is that since there are not 
enough dark matters in outer space, the neutrinos from SN1987A propagated 
at the universal limiting speed $C$ without interacting with dark matters. 
On the other hand, since it is expected that there are sufficient dark matters 
on the earth, the interaction between the photons and dark matters reduces 
the speed of light on the earth to the smaller observed velocity $c(R_\oplus)$ 
from the larger universal limiting speed $C$. Accordingly, by the difference
of existence of dark matters, the OPERA neutrinos travel on the earth at the speed 
less than the universal limiting speed $C$ while the SN1987A ones propagate 
in the interstellar space at the velocity almost equal to the universal limiting 
speed $C$.

\subsection{Vector field}
 
Now we wish to consider the case of a new gauge field denoted as $B_\mu$,
which should not be confused with the gauge field $A_\mu$ describing the conventional
photons. A line of argument in this subsection is similar to the case of the scalar 
field $\Pi$ treated in the previous subsection.

The starting Lagrangian density is defined as 
%**   V-Lag %%%%%%%%%%%%%%%%%%%%%%%%%%%%%%%%%%%%%%%%%%%%%%%%%%%%%%%%%
\begin{eqnarray}
{\cal{L}} &=& - \frac{1}{4} F_{\mu\nu} F^{\mu\nu} 
+ \frac{1}{2 M_*^4} G^{\nu\alpha} G^\beta \ _\alpha  F_{\mu\nu} F^\mu \ _\beta
- \frac{1}{4 M_*^8} G^{\mu\rho} G^\alpha \ _\rho G^{\nu\sigma} G^\beta \ _\sigma  
F_{\mu\nu} F_{\alpha\beta} \nonumber\\
&-& \frac{1}{4} G_{\mu\nu} G^{\mu\nu} - B_\mu J^\mu     \nonumber\\
&=& - \frac{1}{4} (\eta^{\mu\alpha} - \frac{1}{M_*^4} G^{\mu\rho} G^\alpha \ _\rho)
(\eta^{\nu\beta} - \frac{1}{M_*^4} G^{\nu\sigma} G^\beta \ _\sigma) F_{\mu\nu} F_{\alpha\beta}
\nonumber\\
&-& \frac{1}{4} G_{\mu\nu} G^{\mu\nu} - B_\mu J^\mu,
\label{V-Lag}
\end{eqnarray}
%%%%%%%%%%%%%%%%%%%%%%%%%%%%%%%%%%%%%%%%%%%%%%%%%%%%%%%%%%%%%%%%%%%
where the field strength $G_{\mu\nu}$ of the new gauge field $B_\mu$ is defined as
$G_{\mu\nu} = \partial_\mu B_\nu - \partial_\nu B_\mu$, and $B_\mu$ couples
to the electro-magnetic current $J^\mu$ of the earth.
The second expression in the Lagrangian density (\ref{V-Lag}) urges us 
to take an effective metric
%**   V-metric %%%%%%%%%%%%%%%%%%%%%%%%%%%%%%%%%%%%%%%%%%%%%%%%%%%%%%%%%
\begin{eqnarray}
g^{\mu\nu}_{(A)} = \eta^{\mu\nu} - \frac{1}{M_*^4} G^{\mu\rho} G^\nu \ _\rho, 
\label{V-metric}
\end{eqnarray}
%%%%%%%%%%%%%%%%%%%%%%%%%%%%%%%%%%%%%%%%%%%%%%%%%%%%%%%%%%%%%%%%%%%
on which the photon propagates. 

Taking variation of the scalar field $B_\mu$ yields the equation of motion
%**   V-Eq %%%%%%%%%%%%%%%%%%%%%%%%%%%%%%%%%%%%%%%%%%%%%%%%%%%%%%%%%
\begin{eqnarray}
\partial_\mu G^{\mu\nu} &=& J^\nu 
- \frac{1}{M_*^4} \partial_\alpha \left[ ( G^{\beta\alpha} F_\mu \ ^\nu
- G^{\beta\nu} F_\mu \ ^\alpha ) F^\mu \ _\beta \right]
\nonumber\\
&+& \frac{1}{M_*^8} \partial_\alpha \left[ ( G^{\mu\alpha} F^\nu \ _\rho
- G^{\mu\nu} F^\alpha \ _\rho ) G^{\rho\sigma} G^\beta \ _\sigma F_{\mu\beta} \right]. 
\label{V-Eq}
\end{eqnarray}
%%%%%%%%%%%%%%%%%%%%%%%%%%%%%%%%%%%%%%%%%%%%%%%%%%%%%%%%%%%%%%%%%%%
In the linearized level, Eq. (\ref{V-Eq}) reduces to the form
%**   V-L-Eq %%%%%%%%%%%%%%%%%%%%%%%%%%%%%%%%%%%%%%%%%%%%%%%%%%%%%%%%%
\begin{eqnarray}
\Box B_\mu = J_\mu, 
\label{V-L-Eq}
\end{eqnarray}
%%%%%%%%%%%%%%%%%%%%%%%%%%%%%%%%%%%%%%%%%%%%%%%%%%%%%%%%%%%%%%%%%%%
where we have chosen the Lorentz gauge $\partial_\mu B^\mu = 0$.

Now we assume that the electromagnetic current has a static, 
spherically symmetric magnetic source 
%**   J %%%%%%%%%%%%%%%%%%%%%%%%%%%%%%%%%%%%%%%%%%%%%%%%%%%%%%%%%
\begin{eqnarray}
J_\theta = 4 \pi \alpha \delta^3(r), 
\label{J}
\end{eqnarray}
%%%%%%%%%%%%%%%%%%%%%%%%%%%%%%%%%%%%%%%%%%%%%%%%%%%%%%%%%%%%%%%%%%%
where $\alpha$ is the magnetic dipole moment of the earth.
Then, we obtain the solution to the linearized equation of motion (\ref{V-L-Eq})
%**   B %%%%%%%%%%%%%%%%%%%%%%%%%%%%%%%%%%%%%%%%%%%%%%%%%%%%%%%%%
\begin{eqnarray}
B_\theta = - \frac{\alpha}{r}, 
\label{B}
\end{eqnarray}
%%%%%%%%%%%%%%%%%%%%%%%%%%%%%%%%%%%%%%%%%%%%%%%%%%%%%%%%%%%%%%%%%%%
from which we have the non-vanishing magnetic field
%**   G %%%%%%%%%%%%%%%%%%%%%%%%%%%%%%%%%%%%%%%%%%%%%%%%%%%%%%%%%
\begin{eqnarray}
G_{r \theta} = \frac{\alpha}{r^2}. 
\label{G}
\end{eqnarray}
%%%%%%%%%%%%%%%%%%%%%%%%%%%%%%%%%%%%%%%%%%%%%%%%%%%%%%%%%%%%%%%%%%%

As in the scalar field, it is straightforward to derive an effective
metric by using Eq's. (\ref{V-metric}) and (\ref{G}). Then, the line element 
reads
%**   Line-V %%%%%%%%%%%%%%%%%%%%%%%%%%%%%%%%%%%%%%%%%%%%%%%%%%%%%%%%%
\begin{eqnarray}
ds^2 = - dt^2 + \frac{1}{1 - \frac{1}{M_*^4} \frac{\alpha^2}{r^6}} 
( dr^2 + r^2 d \theta^2) + r^2 \sin^2 \theta d \varphi^2.
\label{Line-V}
\end{eqnarray}
%%%%%%%%%%%%%%%%%%%%%%%%%%%%%%%%%%%%%%%%%%%%%%%%%%%%%%%%%%%%%%%%%%%
Using the spherical symmetry, we take $\varphi = 0$. Then, we have
%**   Line-V 2 %%%%%%%%%%%%%%%%%%%%%%%%%%%%%%%%%%%%%%%%%%%%%%%%%%%%%%%%%
\begin{eqnarray}
ds^2|_{\varphi = 0} &=& - dt^2 + \frac{1}{1 - \frac{1}{M_*^4} \frac{\alpha^2}{r^6}} 
( dr^2 + r^2 d \theta^2)  \nonumber\\
&=& - dt^2 + \frac{1}{1 - \frac{1}{M_*^4} \frac{\alpha^2}{r^6}} ( dx^2 + dz^2),
\label{Line-V 2}
\end{eqnarray}
%%%%%%%%%%%%%%%%%%%%%%%%%%%%%%%%%%%%%%%%%%%%%%%%%%%%%%%%%%%%%%%%%%%
from which we can read out an effective local velocity of a photon
%**   V-velocity %%%%%%%%%%%%%%%%%%%%%%%%%%%%%%%%%%%%%%%%%%%%%%%%%%%%%%%%%
\begin{eqnarray}
c(r) &\equiv& C \sqrt{ \left(\frac{d x}{d t}\right)^2 +  
\left(\frac{d z}{d t}\right)^2 }  \nonumber\\
&=& C \sqrt{ 1 - \frac{1}{M_*^4} \frac{\alpha^2}{r^6} }.
\label{V-velocity}
\end{eqnarray}
%%%%%%%%%%%%%%%%%%%%%%%%%%%%%%%%%%%%%%%%%%%%%%%%%%%%%%%%%%%%%%%%%%%
This observed speed of light also satisfies the relation (\ref{S-velocity 3}),
and the difference between the universal limiting speed $C$ and the observed 
one $c(r)$ becomes positive-definite as in the scalar field. 

From the results of OPERA and SN1987A, the dimensionless quantity $\beta$ must take the following
value
%**   V-Beta %%%%%%%%%%%%%%%%%%%%%%%%%%%%%%%%%%%%%%%%%%%%%%%%%%%%%%%%%
\begin{eqnarray}
\beta(R_\oplus) \equiv  \frac{v_\nu - c(R_\oplus)}{c(R_\oplus)} 
\approx \frac{1}{2 M_*^4} \frac{\alpha^2}{R_\oplus^6}
\approx 2.5 \times 10^{-5}.
\label{V-Beta}
\end{eqnarray}
%%%%%%%%%%%%%%%%%%%%%%%%%%%%%%%%%%%%%%%%%%%%%%%%%%%%%%%%%%%%%%%%%%%
Moreover, since the magnetic field on the earth is about $0.5 \times 10^{-4}$
Tesla, we have a relation
%**   Tesla %%%%%%%%%%%%%%%%%%%%%%%%%%%%%%%%%%%%%%%%%%%%%%%%%%%%%%%%%
\begin{eqnarray}
\frac{\alpha}{R_\oplus^3}
= 0.5 \times 10^{-4}.
\label{Tesla}
\end{eqnarray}
%%%%%%%%%%%%%%%%%%%%%%%%%%%%%%%%%%%%%%%%%%%%%%%%%%%%%%%%%%%%%%%%%%%
Recovering dimensional factors, it turns out that the mass scale is described as
%**   V-Scale %%%%%%%%%%%%%%%%%%%%%%%%%%%%%%%%%%%%%%%%%%%%%%%%%%%%%%%%%
\begin{eqnarray}
M_*^4 = 0.2 \times 10^5 \frac{\hbar^3}{\mu_0 c^5}
\left(\frac{\alpha}{R_\oplus^3}\right)^2,
\label{V-Scale}
\end{eqnarray}
%%%%%%%%%%%%%%%%%%%%%%%%%%%%%%%%%%%%%%%%%%%%%%%%%%%%%%%%%%%%%%%%%%%
where $\mu_0$ denotes magnetic permeability of the vacuum on the earth. This
expression gives us the mass scale in this theory
%**   V-Scale 2 %%%%%%%%%%%%%%%%%%%%%%%%%%%%%%%%%%%%%%%%%%%%%%%%%%%%%%%%%
\begin{eqnarray}
M_* \approx  1 eV.
\label{V-Scale 2}
\end{eqnarray}
%%%%%%%%%%%%%%%%%%%%%%%%%%%%%%%%%%%%%%%%%%%%%%%%%%%%%%%%%%%%%%%%%%%
Such a low mass scale has also appeared in our previous work \cite{Oda1}
where a superluminal neutrino has been obtained by coupling the neutrino field
to a new gauge field. However, a physical interpretation of the two low mass scales 
might be different since in the present model there are two gauge fields $B_\mu$ 
and $A_\mu$ whereas in the previous model there is only one gauge field $A_\mu$
which could be identified with the conventional photon field.

\subsection{Tensor field}

A symmetric tensor field was originally dealt with in Ref. \cite{Dvali}
in order to make neutrinos superluminal to explain the OPERA results.
In this subsection, we make use of the symmetric tensor field to obtain the
observed speed of light smaller than the universal limiting speed. We will again
follow a similar path of thought to the cases of scalar and vector fields.

The Lagrangian density consists of three parts which are part of the gauge
field and its coupling with the tensor, that of massive gravity of Fierz-Pauli
type, and that of source as follows:
%**   T-Lag %%%%%%%%%%%%%%%%%%%%%%%%%%%%%%%%%%%%%%%%%%%%%%%%%%%%%%%%%
\begin{eqnarray}
{\cal{L}} &=& - \frac{1}{4} F_{\mu\nu} F^{\mu\nu} 
+ \frac{1}{2 M_*} h^{\nu\alpha} F_{\mu\nu} F^\mu \ _\alpha
- \frac{1}{4 M_*^2} h^{\mu\alpha} h^{\nu\beta} F_{\mu\nu} F_{\alpha\beta}  
\nonumber\\
&+& \frac{1}{2} h_{\mu\nu} {\cal{E}}^{\mu\nu\alpha\beta} h_{\alpha\beta} 
- \frac{m^2}{2} ( h_{\mu\nu}^2 - h^2 ) + \frac{1}{M} h_{\mu\nu} T^{\mu\nu}
\nonumber\\
&=& - \frac{1}{4} (\eta^{\mu\alpha} - \frac{1}{M_*} h^{\mu\alpha})
(\eta^{\nu\beta} - \frac{1}{M_*} h^{\nu\beta}) F_{\mu\nu} F_{\alpha\beta}
\nonumber\\
&+& \frac{1}{2} h_{\mu\nu} {\cal{E}}^{\mu\nu\alpha\beta} h_{\alpha\beta} 
- \frac{m^2}{2} ( h_{\mu\nu}^2 - h^2 ) + \frac{1}{M} h_{\mu\nu} T^{\mu\nu},
\label{T-Lag}
\end{eqnarray}
%%%%%%%%%%%%%%%%%%%%%%%%%%%%%%%%%%%%%%%%%%%%%%%%%%%%%%%%%%%%%%%%%%%
where $h_{\mu\nu}$ is defined as $h_{\mu\nu} \equiv g_{\mu\nu} - \eta_{\mu\nu}$,
and ${\cal{E}}^{\mu\nu\alpha\beta}$ is an operator stemming from the Einstein-Hilbert
term which is defined for a general symmetric tensor $Z_{\alpha\beta} = Z_{\beta\alpha}$ as
%**   Z %%%%%%%%%%%%%%%%%%%%%%%%%%%%%%%%%%%%%%%%%%%%%%%%%%%%%%%%%
\begin{eqnarray}
{\cal{E}}^{\mu\nu\alpha\beta} Z_{\alpha\beta}
= \Box Z^{\mu\nu} - \eta^{\mu\nu} \Box Z - \partial^\mu \partial_\alpha Z^{\nu\alpha}
- \partial^\nu \partial_\alpha Z^{\mu\alpha} + \partial^\mu \partial^\nu Z
+ \eta^{\mu\nu} \partial_\alpha \partial_\beta Z^{\alpha\beta}.
\label{Z}
\end{eqnarray}
%%%%%%%%%%%%%%%%%%%%%%%%%%%%%%%%%%%%%%%%%%%%%%%%%%%%%%%%%%%%%%%%%%%
The second expression in the Lagrangian density (\ref{T-Lag}) leads to 
an effective metric
%**   T-metric %%%%%%%%%%%%%%%%%%%%%%%%%%%%%%%%%%%%%%%%%%%%%%%%%%%%%%%%%
\begin{eqnarray}
g^{\mu\nu}_{(A)} = \eta^{\mu\nu} - \frac{1}{M_*} h^{\mu\nu}.
\label{T-metric}
\end{eqnarray}
%%%%%%%%%%%%%%%%%%%%%%%%%%%%%%%%%%%%%%%%%%%%%%%%%%%%%%%%%%%%%%%%%%%
In contrast to the scalar and vector cases with a factor $\frac{1}{M_*^4}$,
this metric depends on $\frac{1}{M_*}$, which makes the mass scale 
larger around the Planck scale as seen shortly. 

The equation of motion for the tensor field $h_{\mu\nu}$ is of form
%**   T-Eq %%%%%%%%%%%%%%%%%%%%%%%%%%%%%%%%%%%%%%%%%%%%%%%%%%%%%%%%%
\begin{eqnarray}
{\cal{E}}_{\mu\nu} \ ^{\alpha\beta} h_{\alpha\beta} - m^2 ( h_{\mu\nu}
- \eta_{\mu\nu} h )
= - \frac{1}{M} T_{\mu\nu} + \frac{1}{2 M_*} (\eta^{\alpha\beta} 
- \frac{1}{M_*} h^{\alpha\beta}) F_{\mu\alpha} F_{\nu\beta}. 
\label{T-Eq}
\end{eqnarray}
%%%%%%%%%%%%%%%%%%%%%%%%%%%%%%%%%%%%%%%%%%%%%%%%%%%%%%%%%%%%%%%%%%%
In the linearized level, Eq. (\ref{T-Eq}) becomes the form
%**   T-L-Eq %%%%%%%%%%%%%%%%%%%%%%%%%%%%%%%%%%%%%%%%%%%%%%%%%%%%%%%%%
\begin{eqnarray}
{\cal{E}}_{\mu\nu}\ ^{\alpha\beta} h_{\alpha\beta} - m^2 ( h_{\mu\nu}
- \eta_{\mu\nu} h ) = - \frac{1}{M} T_{\mu\nu}. 
\label{T-L-Eq}
\end{eqnarray}
%%%%%%%%%%%%%%%%%%%%%%%%%%%%%%%%%%%%%%%%%%%%%%%%%%%%%%%%%%%%%%%%%%%
Taking the trace of Eq. (\ref{T-L-Eq}), one obtains
%**   T-L-Eq 2 %%%%%%%%%%%%%%%%%%%%%%%%%%%%%%%%%%%%%%%%%%%%%%%%%%%%%%%%%
\begin{eqnarray}
\Box h - \partial_\alpha \partial_\beta h^{\alpha\beta} 
= \frac{3}{2} m^2 h + \frac{1}{2 M} T. 
\label{T-L-Eq 2}
\end{eqnarray}
%%%%%%%%%%%%%%%%%%%%%%%%%%%%%%%%%%%%%%%%%%%%%%%%%%%%%%%%%%%%%%%%%%%
Moreover, taking the divergence of Eq. (\ref{T-L-Eq}) produces an equation
%**   T-L-Eq 3 %%%%%%%%%%%%%%%%%%%%%%%%%%%%%%%%%%%%%%%%%%%%%%%%%%%%%%%%%
\begin{eqnarray}
\partial^\mu h_{\mu\nu} - \partial_\nu h = 0. 
\label{T-L-Eq 3}
\end{eqnarray}
%%%%%%%%%%%%%%%%%%%%%%%%%%%%%%%%%%%%%%%%%%%%%%%%%%%%%%%%%%%%%%%%%%%
Together with Eq. (\ref{T-L-Eq 2}) and Eq. (\ref{T-L-Eq 3}), we can derive
an equation
%**   T-L-Eq 4 %%%%%%%%%%%%%%%%%%%%%%%%%%%%%%%%%%%%%%%%%%%%%%%%%%%%%%%%%
\begin{eqnarray}
h = - \frac{1}{3 m^2 M} T. 
\label{T-L-Eq 4}
\end{eqnarray}
%%%%%%%%%%%%%%%%%%%%%%%%%%%%%%%%%%%%%%%%%%%%%%%%%%%%%%%%%%%%%%%%%%%
Then, Eq.  (\ref{T-L-Eq}) with the help of Eq's. (\ref{T-L-Eq 3}) and (\ref{T-L-Eq 4})
can be cast to the form
%**   T-L-Eq 5 %%%%%%%%%%%%%%%%%%%%%%%%%%%%%%%%%%%%%%%%%%%%%%%%%%%%%%%%%
\begin{eqnarray}
(\Box - m^2) h_{\mu\nu} = - \frac{1}{M} \left[ T_{\mu\nu} - \frac{1}{3}
(\eta_{\mu\nu} - \frac{1}{m^2} \partial_\mu \partial_\nu) T \right]. 
\label{T-L-Eq 5}
\end{eqnarray}
%%%%%%%%%%%%%%%%%%%%%%%%%%%%%%%%%%%%%%%%%%%%%%%%%%%%%%%%%%%%%%%%%%%

Since we take a non-relativistic, static and spherically symmetric source (\ref{S-T}) 
and the long Compton wave-length approximation as in the scalar field, 
we finally obtain the equation of motion for $h_{\mu\nu}$
%**   T-L-Eq 6 %%%%%%%%%%%%%%%%%%%%%%%%%%%%%%%%%%%%%%%%%%%%%%%%%%%%%%%%%
\begin{eqnarray}
\Delta h_{00} = - \frac{2}{3 M} M_\oplus \delta^3(r), \quad
\Delta h_{ij} = - \frac{1}{3 M} M_\oplus \delta^3(r) \delta_{ij},
\label{T-L-Eq 6}
\end{eqnarray}
%%%%%%%%%%%%%%%%%%%%%%%%%%%%%%%%%%%%%%%%%%%%%%%%%%%%%%%%%%%%%%%%%%%
where indices $i, j$ indicate spatial ones $i = 1, 2, 3$ and the contribution 
proportional to total derivative is neglected owing to conservation of
the electro-magnetic current. Using the
Poisson equation $\Delta \frac{1}{r} = - 4 \pi \delta^3(r)$, Eq. (\ref{T-L-Eq 6})
can be solved to the form
%**   h-sol %%%%%%%%%%%%%%%%%%%%%%%%%%%%%%%%%%%%%%%%%%%%%%%%%%%%%%%%%
\begin{eqnarray}
h_{00} = \frac{1}{6 \pi} \frac{M_\oplus}{M} \frac{1}{r}, \quad
h_{ij} = \frac{1}{12 \pi} \frac{M_\oplus}{M} \frac{1}{r} \delta_{ij}.
\label{h-sol}
\end{eqnarray}
%%%%%%%%%%%%%%%%%%%%%%%%%%%%%%%%%%%%%%%%%%%%%%%%%%%%%%%%%%%%%%%%%%%

As before, it is easy to calculate the effective metric (\ref{T-metric}) by using 
Eq. (\ref{h-sol}). Then, taking the inverse of the effective metric, 
the line element is found to be
%**   Line-T %%%%%%%%%%%%%%%%%%%%%%%%%%%%%%%%%%%%%%%%%%%%%%%%%%%%%%%%%
\begin{eqnarray}
ds^2 = - \left( 1 - \frac{1}{6 \pi} \frac{M_\oplus}{M_* M} \frac{1}{r} \right) dt^2 
+ \left( 1 + \frac{1}{12 \pi} \frac{M_\oplus}{M_* M} \frac{1}{r} \right) 
\delta_{ij} d x^i d x^j. 
\label{Line-T}
\end{eqnarray}
%%%%%%%%%%%%%%%%%%%%%%%%%%%%%%%%%%%%%%%%%%%%%%%%%%%%%%%%%%%%%%%%%%%
Since the trajectory of light is null, $ds^2 = 0$, the effective local velocity 
of a photon reads
%**   T-velocity %%%%%%%%%%%%%%%%%%%%%%%%%%%%%%%%%%%%%%%%%%%%%%%%%%%%%%%%%
\begin{eqnarray}
c(r) &\equiv& C \sqrt{ \delta_{ij} \frac{dx^i}{dt} \frac{dx^j}{dt} }  \nonumber\\
&=& C \sqrt{ \frac{ 1 - \frac{1}{6 \pi} \frac{M_\oplus}{M_* M} \frac{1}{r} }
{ 1 + \frac{1}{12 \pi} \frac{M_\oplus}{M_* M} \frac{1}{r} } }
\nonumber\\
&\approx& C \left(1 - \frac{1}{8 \pi} \frac{M_\oplus}{M_* M} \frac{1}{r} \right).
\label{T-velocity}
\end{eqnarray}
%%%%%%%%%%%%%%%%%%%%%%%%%%%%%%%%%%%%%%%%%%%%%%%%%%%%%%%%%%%%%%%%%%%
This observed speed of light also satisfies the relation (\ref{S-velocity 3}),
and the difference between $C$ and $c(r)$ becomes
positive-definite as in the both scalar and vector fields. 

Again the results of OPERA and SN1987A require the dimensionless quantity $\beta$ to
take the following value
%**   T-Beta %%%%%%%%%%%%%%%%%%%%%%%%%%%%%%%%%%%%%%%%%%%%%%%%%%%%%%%%%
\begin{eqnarray}
\beta(R_\oplus) \equiv  \frac{v_\nu - c(R_\oplus)}{c(R_\oplus)} 
\approx \frac{1}{8 \pi} \frac{M_\oplus}{M_* M} \frac{1}{R_\oplus}
\approx 2.5 \times 10^{-5}.
\label{T-Beta}
\end{eqnarray}
%%%%%%%%%%%%%%%%%%%%%%%%%%%%%%%%%%%%%%%%%%%%%%%%%%%%%%%%%%%%%%%%%%%

Using this expression and recovering dimensional factors, it turns out that 
the two mass scales are constrained to be
%**   T-Scale %%%%%%%%%%%%%%%%%%%%%%%%%%%%%%%%%%%%%%%%%%%%%%%%%%%%%%%%%
\begin{eqnarray}
M_* M  = \frac{1}{40 \pi} \times 10^5 M_{Pl}^2 \frac{R_{\oplus SS}}{R_\oplus} 
\approx 10^{-6} M_{Pl}^2.
\label{T-Scale}
\end{eqnarray}
%%%%%%%%%%%%%%%%%%%%%%%%%%%%%%%%%%%%%%%%%%%%%%%%%%%%%%%%%%%%%%%%%%%
This result is very similar to that obtained by Dvali and Vikman in Ref. \cite{Dvali}
so that we think that the present model passes various phenomenological
constraints investigated in \cite{Dvali, Iorio}. For example, the absence of any
observable long-range fifth force of gravity-type and the maximal violation
of equivalence principle imply $M \approx 10^2 M_{Pl}$ and $M_* \approx 
10^{-8} M_{Pl}$. The latter mass scale is also consistent with the correction
to the cooling rate of stars because of production of $h_{\mu\nu}$.
Finally, we remark that in contrast to the theory in \cite{Dvali}, in our theory 
there is no sign asymmetry of the couplings so our theory could have a consistent 
UV-completion.

The last, but not least, let us comment on the stability of a photon which
is known to be stable \cite{Nakamura}. In this article, we have introduced 
specific non-renormalizable interactions to obtain the observed speed of light
from the universal limiting speed. These interaction terms have a 
possibility of triggering the spontaneous decay of a photon, thus losing
its stability. However, it turns out that the decay process occurs at 
the loop levels (in cases of the both scalar and tensor models, it occurs
from the two loop level while in the vector model, at the one loop level), 
and it is greatly suppressed by the mass scale $M_*$. For instance,
since in the scalar model, the amplitude $\gamma + \gamma \rightarrow \Pi + \Pi$
is schematically described as $\left(\frac{1}{M_*^4}\right)^3 \left(\int^\Lambda
d^4 p \frac{p^6}{p^4}\right)^2 \approx \left(\frac{\Lambda}{M_*}\right)^{12}$, 
the amplitude is so tiny as long as $\Lambda \ll M_*$, thereby ensuring the stability 
of a photon in this model. In a similar way, it is easily shown that a photon is
stable for  $\Lambda \ll M_*$ in all the models.

%%%%%%%%%%%%%%%%%%%%%%%%%%%%%%%%%%%%%%%%%%%%%%%%%%%%%%%%%%%%%%%%%%%%%
%%%%%%%%%%%%%%%%%%%%%%%%%%%%%%   SEC  4    %%%%%%%%%%%%%%%%%%%%%%%%%%
%%%%%%%%%%%%%%%%%%%%%%%%%%%%%%%%%%%%%%%%%%%%%%%%%%%%%%%%%%%%%%%%%%%%%
\section{Discussion}

In this article, we have explained our resolution to the Cohen-Glashow effect
associated with the OPERA superluminal neutrinos in detail. After all,
our resolution is equivalent to saying that the OPERA neutrinos are not
actually superluminal but subluminal since they travel at the speed 
slower than a universal limiting speed. 

Relating to our resolution, we have also pointed out that the principle of 
invariant speed of light in special relativity can be replaced with the principle 
of a universal limiting speed. This universal limiting speed is the maximum velocity 
of elementary particles and information, and the observed speed of light must be 
equal to or smaller than the universal limiting speed where the equality holds 
when the coupling of a photon with dark matters is switched off. 
However, since our earth is surrounded by a cloud 
of dark matters and cannot escape from the influence, the observed speed 
of light must be always less than the universal limiting speed. As emphasied in this
article, the causal structure of all the events should be defined through not
the observed speed of light on the earth but the universal limiting speed. Furthermore,
this universal limiting speed is nothing but the velocity appearing in various
formulae of special relativity and electrodynamics such as the Lorentz transformation.

Our resolution to the Cohen-Glashow effect is on the kinematical grounds and thus strictly 
forbids the OPERA neutrinos to emit a pair of electron and positron via the bremsstrahlung 
effect. It is then natural to ask ourselves if our resolution also provides a resolution 
to the other problems associated with the OPERA results. In particular, it is now known that
the other challenging theoretical issue lies in pion decay process  \cite{Gonzalez, Bi, Cowsik}
where it was found that the decay of charged pion $\pi^+ \rightarrow \mu^+ + \nu_\mu$,
which is nothing but the neutrino production process in the OPERA experiment,
becomes kinematically forbidden for $E_\nu > 5 GeV$, which is obviously inconsistent
with the OPERA results. It is worth stressing that our resolution also provides a solution 
to this problem since this decay process is automatically prohibitted whenever the OPERA 
neutrinos are subluminal.

Nevertheless, there remain some issues to be clarified in future. In particular, we should
investigate various phenomenological implications of the present models in more detail. 
For instance, the couplings between the new bosonic fields and photons would yield
observable effects on the bending and red-shift of light. Furthermore, we should look
for astrophysical and cosmological constraints coming from CMB and supernova Ia. 
Dark matters in the universe are known to be so dilute but the propagation over
cosmological distances might give rise to an accumulated contribution even for a tiny
coupling. We wish to take into consideration these problems in the next publication.

%%%%%%%%%%%%%%%%%%%%%%%%%%%%%%%%%%%%%%%%%%%%%%%%%%%%%%%%%%%%%%%%%%
%%%%%%%%%%%%%%%%%%%%%%%% Acknowledgements %%%%%%%%%%%%%%%%%%%%%%%%%%%%%
%%%%%%%%%%%%%%%%%%%%%%%%%%%%%%%%%%%%%%%%%%%%%%%%%%%%%%%%%%%%%%%%%%
\begin{flushleft}
{\bf Acknowledgements}
\end{flushleft}

We wish to thank Prof. N. Nakanishi for valuable discussions.
This work is supported in part by the Grant-in-Aid for Scientific 
Research (C) No. 22540287 from the Japan Ministry of Education, Culture, 
Sports, Science and Technology.

%%%%%%%%%%%%%%%%%%%%%%%% reference %%%%%%%%%%%%%%%%%%%%%%%%%%%%%%%
%%%%%%%%%%%%%%%%%%%%%%%%%%%%%%%%%%%%%%%%%%%%%%%%%%%%%%%%%%%%%%%%%%

\end{document}